\documentclass[sigconf,screen]{acmart}
\AtBeginDocument{%
  \providecommand\BibTeX{{%
    \normalfont B\kern-0.5em{\scshape i\kern-0.25em b}\kern-0.8em\TeX}}}

\setcopyright{acmcopyright}
\copyrightyear{2018}
\acmYear{2018}
\acmDOI{XXXXXXX.XXXXXXX}

\acmConference[Conference acronym 'XX]{Make sure to enter the correct
  conference title from your rights confirmation emai}{June 03--05,
  2018}{Woodstock, NY}
\acmBooktitle{Woodstock '18: ACM Symposium on Neural Gaze Detection,
 June 03--05, 2018, Woodstock, NY} 
\acmPrice{15.00}
\acmISBN{978-1-4503-XXXX-X/18/06}

\usepackage{algorithm}
\usepackage{algpseudocode}
\usepackage{xcolor}
\hypersetup{
    colorlinks=true,
    citecolor=blue,
    linkcolor=red,
    urlcolor=orange,
}
\usepackage{float}
\newfloat{Prompt}{h}{lop}
\floatname{Prompt}{Prompt}

\usepackage{listings}
\begin{document}

\title{Towards Digital Nature: Bridging the Gap between Turing Machine Objects and Linguistic Objects in LLMMs for Universal Interaction of Object-Oriented Descriptions}

\author{Yoichi Ochiai}
\email{wizard@slis.tsukuba.ac.jp}
\authornotemark[1]
\orcid{0000-0002-4690-5724}
\author{Naruya Kondo}
\author{Tatsuki Fushimi}
\affiliation{%
  \institution{Research and Development Center for Digital Nature, University of Tsukuba}
  \streetaddress{1-2, Kasuga}
  \city{Tsukuba}
  \state{Ibaraki}
  \country{Japan}
  \postcode{305-8550}
}

\renewcommand{\shortauthors}{Trovato and Tobin, et al.}

\begin{abstract}
In this paper, we propose a novel approach to establish a connection between linguistic objects and classes in Large Language Model Machines (LLMMs) such as GPT3.5 and GPT4, and their counterparts in high-level programming languages like Python. Our goal is to promote the development of Digital Nature—a worldview where digital and physical realities are seamlessly intertwined and can be easily manipulated by computational means.
To achieve this, we exploit the inherent abstraction capabilities of LLMMs to build a bridge between human perception of the real world and the computational processes that mimic it. This approach enables ambiguous class definitions and interactions between objects to be realized in programming and ubiquitous computing scenarios. By doing so, we aim to facilitate seamless interaction between Turing Machine objects and Linguistic Objects, paving the way for universally accessible object-oriented descriptions.
We demonstrate a method for automatically transforming real-world objects and their corresponding simulations into language-simulable worlds using LLMMs, thus advancing the digital twin concept. This process can then be extended to high-level programming languages, making the implementation of these simulations more accessible and practical.
In summary, our research introduces a groundbreaking approach to connect linguistic objects in LLMMs with high-level programming languages, allowing for the efficient implementation of real-world simulations. This ultimately contributes to the realization of Digital Nature, where digital and physical worlds are interconnected, and objects and simulations can be effortlessly manipulated through computational means.
Demos are available at \url{https://codesandbox.io/s/alos-simulator-mk0k2t}.
\end{abstract}



\begin{teaserfigure}
  \includegraphics[width=\textwidth]{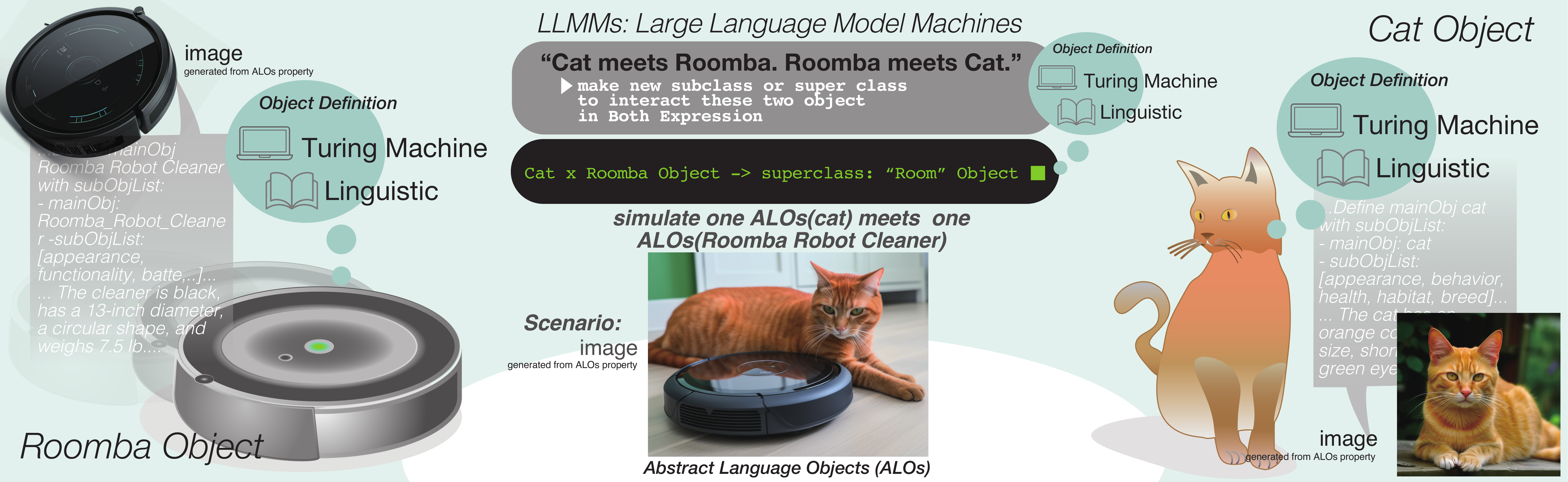}
  \caption{The motivation for this study: In this study, we address the challenge faced by programmers in manually generating comprehensive classes to describe diverse object interactions. Using a Roomba robot vacuum cleaner and a cat as examples, we aim to automatically define their interactions using Large Language Models (LLMs) and generate Turing Machine Objects when needed. We propose connecting Turing Machine Objects and Linguistic Objects through abstract language objects (ALOs), derived from the lexical meanings in LLMs.}
  \label{fig:teaser}
\end{teaserfigure}

\received{20 February 2007}
\received[revised]{12 March 2009}
\received[accepted]{5 June 2009}

\maketitle

\section{Introduction}
The world we inhabit is a rich and intricate mosaic of both living and inanimate objects, all intricately interconnected in countless ways. Throughout human history, we have sought to comprehend and manipulate these objects to satisfy our needs, desires, and inquisitive nature. One of the most fundamental ways humans have interacted with the world is through the act of naming, categorizing, and describing the objects and phenomena that we observe. We can refer to these conceptual entities as "linguistic objects," (LO) which are represented and conveyed through human language.
\begin{figure*}[t]
\includegraphics[width=0.85\textwidth]{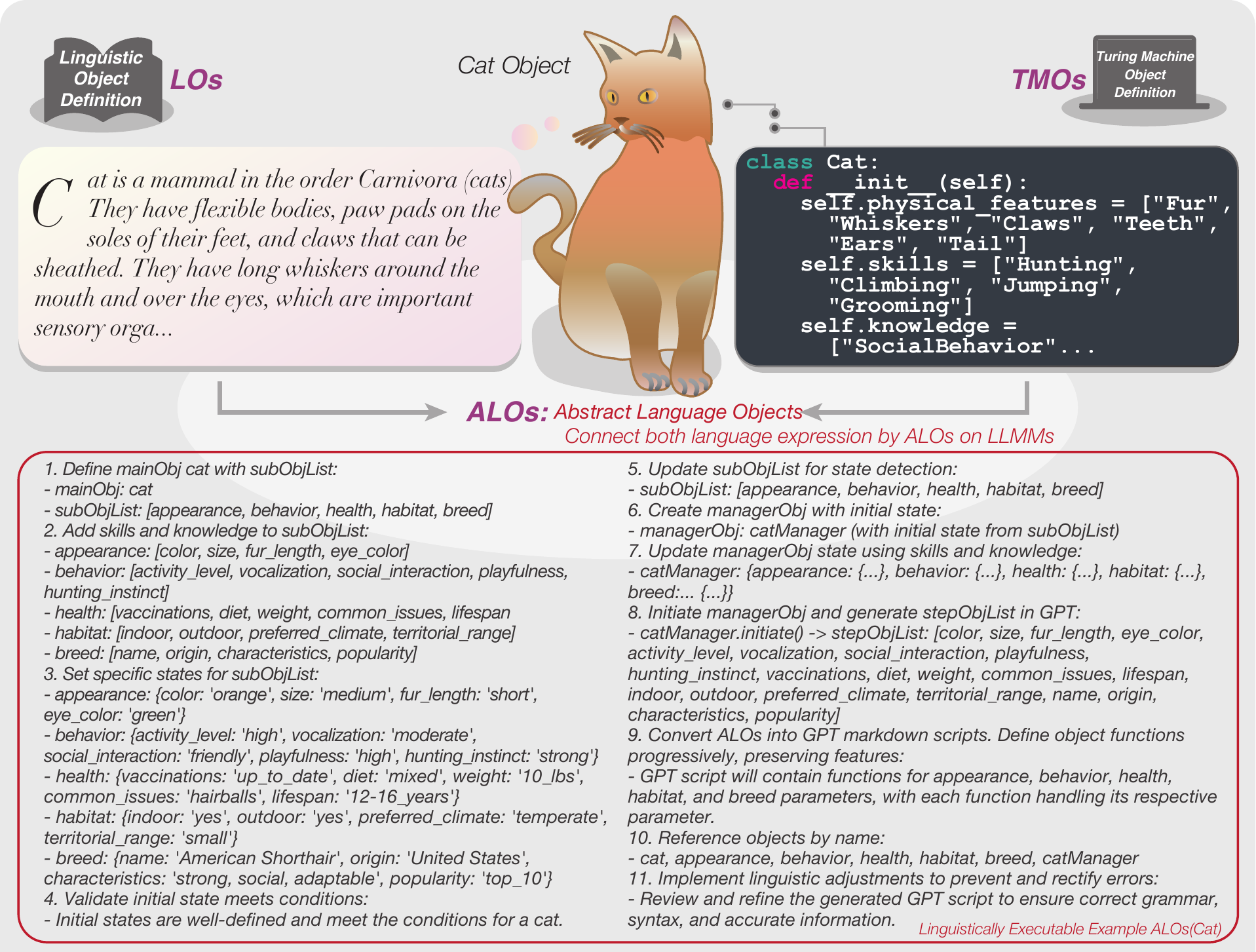}
\caption{Comparison between Linguistic Objects (LOs), Turing Machine Objects (TMOs) and Abstract Language Objects (ALOs): We show the difference between Linguistic Objects (LOs), which have a lexical meaning, Turing Machine Objects (TMOs), which are written in a programming language, and Abstract Language Objects (ALOs), which are defined in this paper. These ALOs can actually be entered into GPT-4 as prompts to run the simulation linguistically on GPT.}
\centering
\label{lo_vs_tmo}
\end{figure*}
The process of naming and organizing these linguistic objects not only facilitates communication but also helps structure our understanding of reality, empowering us to engage with it more effectively. The collection of these linguistic objects is compiled in what we know as a dictionary, serving as a repository for our shared knowledge and a testament to the ever-evolving tapestry of human language and experience.

The evolution of computing, particularly object-oriented programming languages (OOP), has revolutionized our ability to represent and interact with real-world objects~\cite{Nygaard1978, Kay1996}. OOP's power in modeling and simulating complex systems is undeniable, yet the translation between real-world and digital objects remains a challenging, labor-intensive, and knowledge-intensive task that demands significant human effort and expertise. Within OOP languages such as Python and Java, we can define constructs known as Turing Machine Objects (TMOs), which encapsulate data and procedures to facilitate abstraction, encapsulation, inheritance, and polymorphism. However, in order to achieve seamless interaction and manipulation between the digital and physical worlds, further research is needed to bridge the gap between these TMOs and human perception via LO.

In the Age of Enlightenment, Descartes posited that the universe could be understood as a vast machine, its intricate workings comprehensible through the language of mathematics and logic (Descartes, 1641). Today, we stand at the threshold of a new age - one where the digital and physical realities converge, allowing humans to engage with and manipulate the world through computational means. This emerging paradigm, which we term Digital Nature, has the potential to revolutionize human interaction with our surroundings and reshape the way we perceive and engage with reality.

In recent years, advances in machine learning and natural language processing have given rise to large language models (LLM) such as GPT3.5 and GPT-4~\cite{OpenAI2023, Bommarito2022} , which have demonstrated remarkable capabilities in understanding and generating human-like text. These models, trained on vast amounts of textual data, possess an impressive ability to learn and infer knowledge about the world, as well as to generate creative and coherent descriptions of objects and their behaviors. Their possibilities are now expanding to realize a LLM to act as machines (large language model machines - LLMMs)\cite{Giannou2023}. 

In this paper, we present Abstract Language Objects (ALOs) refer to a novel way of bridging LOs and TMOs in the context of OOP and language simulation. ALOs are designed to facilitate seamless interaction between human-perceived real-world entities and their computational counterparts by employing the abstraction capabilities of LLMMs and TMOs. By systematically defining, validating, and updating the states and relationships of these objects, ALOs enable the efficient implementation and manipulation of real-world simulations and contribute to the development of Digital Nature. 

This research is situated at the intersection of computer-human interaction, linguistics, and artificial intelligence. By exploring the potential of LLMMs to create language-simulable worlds, we contribute to the growing body of work in the field of human-computer interaction that seeks to enhance our ability to engage with and manipulate digital and physical realities. The implications of this research are far-reaching, touching upon areas such as virtual and augmented reality~\cite{Milgram1994, Azuma1997}, ubiquitous computing ~\cite{weiser1991}, the digital twin concept~\cite{Barricelli2019}, and the development of natural language interfaces for complex systems~\cite{Androutsopoulos1995}.

The remainder of this paper is structured as follows. In Section 2, we provide a brief overview of related work in the fields of HCI. Section 3 introduces our novel approach with ALOs. In Section 4, we present a case study that demonstrates the practical implementation of our approach in GPT-4. Finally, Section 5 discusses the implications of our research for human-computer interaction and draw backs, and Section 6 concludes the paper and suggests directions for future research.

%

\section{Related Work}


In recent years, the advancement of LLMs has led to significant breakthroughs in various applications, including story writing, web design, mobile user interfaces, email writing, robotics, and menu system design. This growing body of research demonstrates the potential of LLMs for creating more dynamic and natural interactions between humans and digital systems.

One area that has experienced a considerable impact from LLMs is HCI. \cite{designspace2022} explores the interplay between HCI and generative models, examining how HCI can impact generative models and how generative models can impact HCI. This research highlights the importance of understanding and enhancing the relationship between these two domains to improve user experiences and achieve more efficient, effective, and enjoyable interactions.

In the field of creative writing, several studies have showcased the capabilities of LLMs in generating narratives and stories. \cite{osone2021buncho}, \cite{chung2022talebrush}, \cite{coenen2021wordcraft}, and \cite{yuan2022wordcraft} have demonstrated the potential of LLMs in story writing, showing how these models can be used to produce engaging and coherent narratives.

Web design is another application that has seen advancements due to LLMs. \cite{kim2022stylette} presents a method for modifying web designs using LLMs, allowing users to easily adapt and customize their websites according to their requirements and preferences.

LLMs have also been employed in the development of conversational interactions on mobile UIs. \cite{wang2022enabling} investigates the use of LLMs in enabling more natural and intuitive conversations between users and mobile devices, thereby enhancing the overall usability of mobile applications.

In the context of professional communication, \cite{goodman2022lampost} explores the application of LLMs in email writing. This study demonstrates the potential of these models in generating well-structured and contextually appropriate emails, thereby streamlining the process of drafting and sending professional correspondence.

Robotics is another area that has benefited from the integration of LLMs. \cite{ahn2022can} presents a method for interpreting and executing robot commands using LLMs, allowing for more effective control and communication between humans and robots.

LLMs have also shown promise in the design of menu systems. \cite{kargaran2023menucraft} explores the use of these models in menu system design, demonstrating their potential for creating more intuitive and user-friendly interfaces.
\begin{figure*}[t]
\includegraphics[width=0.9\textwidth]{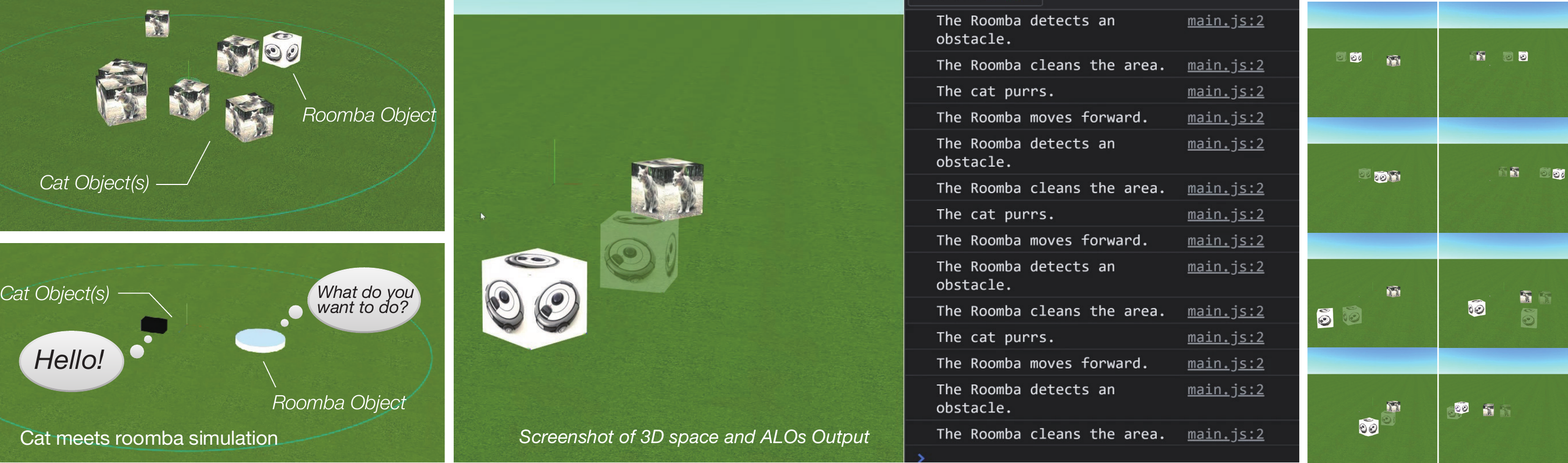}
\caption{ 3D virtual world using Three.js:  (Left)Example of ALOs generation in 3D space (Center) Screenshot of 3D space and ALOs output, (Right) Example Screenshot of 3D objects. Left rows are shown cat object does not avoid Roomba object, Right rows cat object actively interact with Roomba object}
\centering
\label{js_example}
\end{figure*}
Understanding the intricacies of utilizing prompts in large-scale language models is essential for producing more natural-sounding text. \cite{dang2023choice} investigates how users can effectively employ prompts to achieve this goal, providing valuable insights into optimizing the interaction between users and LLMs.

In addition to these applications, LLMs have been employed in educational and assistive contexts. For instance, \cite{duolingo} discusses the use of GPT-4 in supporting Duolingo, a platform for learning foreign languages. This application demonstrates the potential of LLMs in language education, facilitating more effective and engaging learning experiences for users.
Similarly, \cite{bemyeyes} describes the integration of GPT-4 in Be My Eyes, an application that provides assistance for people with low vision. This example illustrates the potential of LLMs in enhancing accessibility and support for individuals with visual impairments.

The growing body of research on LLMs and their applications in various domains highlights the potential of these models for revolutionizing human-computer interaction and creating more natural, dynamic, and accessible digital experiences. By understanding and exploiting the capabilities of LLMs, researchers and practitioners can continue to push the boundaries of what is possible in the digital realm. This line of research could significantly reduce the barrier for non-programmers to create and manipulate digital objects, making these digital ecosystems more accessible.

One potential avenue for research in this area is the development of meta-models and ontologies that can bridge the gap between LLMMs and TMOs, as well as between digital and physical objects. By creating a common framework for describing and manipulating objects in both the digital and physical worlds, researchers can pave the way for the realization of the transformation easily and smoothly between objects used in human-computer interactions.

Nature is a relentless force, while technology falters. The concept of Digital Nature posits a world where digital and physical realities are seamlessly intertwined, allowing for easy manipulation through computational means. This concept has significant implications for the future of information ecosystems and human-computer interactions, as it requires the development of new tools, techniques, and paradigms to enable seamless interaction between humans and machines.


\section{Materials and Methods}

The first step in our approach is the naming and description of real-world objects, which enables the establishment of a connection between the object and its digital representation. This process is inspired by the way humans name new species, items, technologies, etc. To automate this process, we leverage the capabilities of LLMMs, which have been pretrained on vast amounts of textual data and can effectively generate contextually relevant and coherent text based on a given input prompt.

\begin{Prompt}
\caption{System Prompt: ALOs Object Creation}
\begin{algorithmic}[1]
\State Create Abstract Language Objects (ALOs) for {input} using steps 1-11.
\State Define mainObj with subObjList or Skip. Birth of ALOs affects all other ALOs. 
\State Add skills and knowledge to subObjList or Reload.
\State Set specific states for subObjList or Reload.
\State Validate initial state meets conditions or Skip.
\State Update subObjList for state detection or Reload.
\State Create managerObj with initial state or Reload.
\State Update managerObj state using skills and knowledge.
\State Initiate managerObj and generate stepObjList in GPT or Update both suitable to environment.
\State Convert ALOs into GPT markdown scripts. Define object functions progressively, preserving features. 
\State Reference objects by name. Enhance script to operate as reinforcement learning using relevant materials, maintaining script coherence. 
\State Implement linguistic adjustments to prevent and rectify errors.
\end{algorithmic}
\label{alo_creation}
\end{Prompt}

Given an object, LLMMs first extract its features and properties, such as its shape, color, size, and functionality, using the domain knowledge gained through its vast amount of training data. This process can further be aided using computer vision techniques and domain-specific knowledge of the users, to further specify the behavior of ALOs. Then, we feed these features into the LLMM as a natural language prompt, which generates a unique name and description for the object. Using OpenAI's GPT-4 as an example, the platform for defining ALO can be achieved by setting the prompt~\ref{alo_creation} as the system definition of the API (sets the general behavior of the LLM). This linguistic object serves as the foundation for the subsequent steps in our approach, including object interaction and simulation, and integration with high-level programming languages.

Once we have named and described the object, we exploit the abstraction capabilities of LLMMs to build a language-simulable world, where the object can interact with other objects in a semantically meaningful way.

The final component of our approach is the integration of the LLMM-generated linguistic objects and their interactions with high-level programming languages (such as Python or JavaScript), hardware, and neural network platform. This is achieved by automatically transforming the natural language descriptions and interactions generated by the LLMM into corresponding object-oriented code, and prompts using LLMs.

\section{Examples and Case Studies}
In this section, we demonstrate the effectiveness of our approach through several examples and case studies that highlight the potential of connecting LOs in LLMs with high-level programming languages. We present three case studies that showcase the practicality of our approach in different application domains: (1) a smart home environment, (2) an interactive educational simulation, and (3) IoT scenarios. 


\subsection{Case Study 1: a smart home environment: 3D Virtual World}



In this case study, we set out to create a 3D virtual world using Three.js, a JavaScript library that enables 3D graphics in web browsers. Our aim was to develop a simple computer graphics environment with minimal features, such as a ground and sky, but without a physics engine. This allowed us to focus on the integration of ALO Objects and their interactions within the virtual world.

\subsubsection{ALO Generation and Conversion to JavaScript}

The process begins by setting a system prompt to request GPT4 to generate an ALOs in JavaScript (Prompt \ref{alo_js}). Next, we have GPT4 transcribe the generated JavaScript code into a single JavaScript class. The resulting class is then saved as a JavaScript file. In the 3D simulation's main program, we import the class, instantiate it, and register it within the JavaScript animation loop to begin the simulation. Since the main program of the simulator is implemented by humans, manual integration is limited to essential aspects of incorporating the generated class into the simulator.

\subsubsection{3D Model Preparation}
As the proof of concept, we prepared the 3D models for LLM to use, but LLMs often showed capability to generate their own 3D models. Here, we represented all ALOs in the simulator as cubes of the same size for the sake of simplicity. For visual clarity, we used images generated by an image generation model (DALL·E 2) with the ALO's name as the query for the cube's texture.

When generating ALOs, explicitly stating that the 3D library (Three.js) is available in the system prompt (i.e.~prompt ~\ref{alo_js}) consistently generated ALO's JavaScript to include 3D object usage. We modified the generated class definition to allow the use of the pre-prepared 3D objects by passing them as arguments in the class's constructor and methods.

\subsubsection{Error Handling and Code Integration}
If an execution error occurred, we either asked GPT-4 for a solution to fix it. For a minor mistakes, we manually intervened to correct errors. If additional complex implementations were required, we commented out the relevant sections and partially incorporated the generated code. Error handling is a crucial aspect of our implementation, as it ensures the seamless integration of auto-generated code into the 3D simulation. Ongoing research suggest capabilities of LLM to self-correct the codes, and this section has potentials to be automated in the future\footnote{https://yoheinakajima.com/task-driven-autonomous-agent-utilizing-gpt-4-pinecone-and-langchain-for-diverse-applications/}. 

\subsubsection{Results}
The results are as shown in Figure~\ref{js_example} and the supplementary video. LLM was able to follow the system prompt, and generate Javascript codes for creating cat and Roomba ALO. Cat ALO has specific function such as to jump, and to meow, and Roomba ALO was able to move and rotate around its axis. When prompted to interact with each other, Roomba demonstrated capabilities to escape or avoid the cats in the simulated environment.

\begin{Prompt}
\caption{System Prompt: ALOs Object Creation with JavaScript}
\begin{algorithmic}[1]
\State Create or update Abstract Language Objects (ALOs) for {input} using steps 1-10 and write them down in executable javascript + Three.js code (except step 10).
\State Define mainObj with subObjList. Birth of ALOs affects all other ALOs.
\State Add skill list and knowledge list to subObjList.
\State Set a specific state list to subObjList.
\State Validate current state meets specific conditions.
\State Update subObjList for state detection list.
\State Update subObjList for skill execution list.
\State Create managerObj of mainObj with initial state.
\State Update managerObj state using skills and knowledge.
\State Generate specific stepObjList for the current states using skills and knowledge.
\State Convert ALOs into GPT markdown scripts.
\State Define object functions progressively preserving features. When updating, enhance script to operate as reinforcement learning using relevant materials, maintaining script coherence, perform linguistic adjustments to prevent and rectify errors, and write all the codes again. If other instructions come, follow them.
\end{algorithmic}
\label{alo_js}
\end{Prompt}

\begin{Prompt}
\caption{User Prompt: ALOs Object Creation}
\begin{algorithmic}[1]
\State Create ALOs(cat)
\State Convert all definitions for ALOs(cat) into a single class.
\State Define updateCatPerFrame function which will be called by the global animate function every frame.
\State Create ALOs(3D physical world)
\State Convert all definitions for ALOs(3D physics world) into a single class.
\State Define updatePhysicalWroldPerFrame function which will be called by the global animate function every frame.
\State ALOs(cat) lives in a ALOs(3D physical world). Redefine ALOs(cat) to fit this situation from scratch using step 1-10.
\State Convert all definitions for ALOs(cat) above into a single class following to steps 1-10 in order.
\State Define updateCatPerFrame function which will be called by the global animate function every frame.
\State ALOs(roomba) lives in ALOs(3D physical world). Create ALOs(roomba) to fit this situation properly.
\State Convert all definitions for ALOs(roomba) above into a single class following to steps 1-10 in order.
\State Define updateRoombaPerFrame function which will be called by the global animate function every frame.
\State ALOs(cat) meets ALOs(roomba) in ALOs(bounded 3D physical world). Create ALOs(cat meets roomba)
\State Convert all definitions for catMeetsRoomba above into a single class following to steps 1-10 in order.
\State Define updateCatMeetsRoombaPerFrame function which will be called by the global animate function every frame.
\end{algorithmic}
\label{alo_jsuser}
\end{Prompt}



\subsection{Case Study 2: Classroom Simulation}

\subsubsection{Creating a Classroom}

To automatically simulate a classroom language using ALOs, we first create ALOs representing the essential components of a classroom environment. These include ALOs for the classroom itself, students, and a teacher. This process is demonstrated in Prompt~\ref{classroom_user_prompt}, which provides an example of how to define the parameters and interactions for each ALO and Prompt~\ref{alo_creation} was used as the system prompt.

\subsubsection{Generating Visuals using Image Generation Software}
\begin{figure*}[t]
\includegraphics[width=\textwidth]{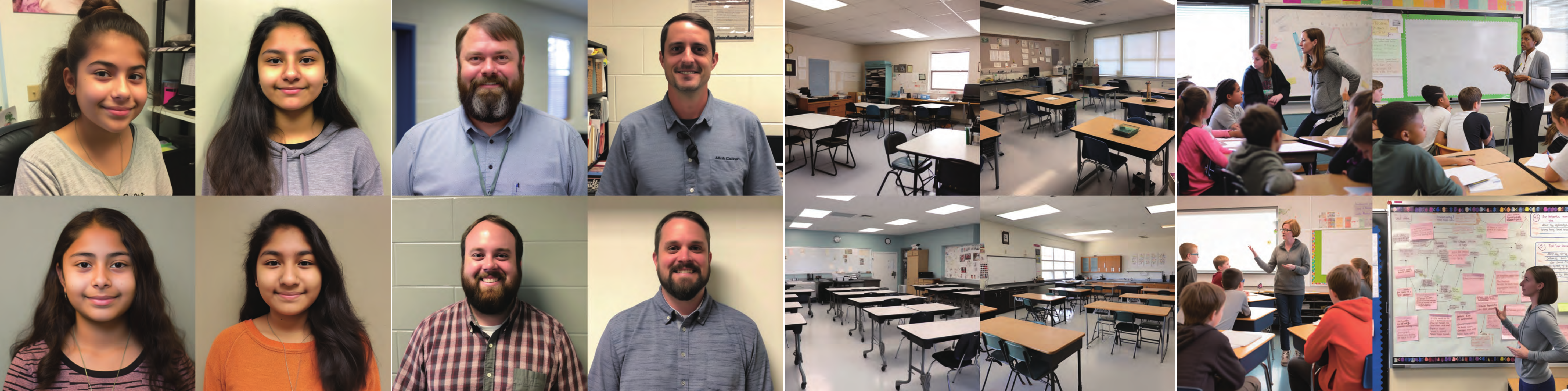}
\caption{Visualization of ALOs(Student), ALOs(Teacher), ALOs(Classroom), and ALOs(classroom).senario in Midjourney V5 by Prompt 5}
\centering
\label{classroom_example}
\end{figure*}

Once the ALOs have been created, we can generate visual representations of these objects by inputting their defined parameters into an image generation software such as Midjouney V5. The results are as shown in Figure~\ref{classroom_example}, and demonstrates that LLMMs can extract information from the defined ALOs to create visualizations of each object. 

This method can be applied to all ALOs; however, certain scenarios (e.g., 4.3) may result in outputs that resemble diagrams rather than scenes due to the lack of visual information in the initial parameters. It is important to note that there is room for improvement regarding the image generation parameter filling process. As demonstrated in 4.3, LLMs often describe the performance specification, leaving parameters related to visuals to be omitted. This can result in less accurate or incomplete visual representations of the ALOs. 

\begin{Prompt}
\caption{User Prompt: ALOs Object Creation for Teacher, Student and Classroom}
\begin{algorithmic}[1]
\State ALOs(classroom) and brainstorm all parameters step-by-step to add and fill.
\State get ALOs(classroom) object and brainstorm to fill subobject parameters and output one ALOs(classroom) object subobject list and parameters in table
\State ALOs(Student) and brainstorm all parameters step-by-step to add and fill.
\State get ALOs(student) object and brainstorm to fill subobject parameters and output one ALOs(student) object subobject list and parameters in table
\State ALOs(Teacher) and brainstorm all parameters step-by-step to add and fill.
\State get ALOs(Teacher) object and brainstorm to fill subobject parameters and output one ALOs(teacher) object subobject list and parameters in table
\State simulate ALOs(teacher) teaches  25 ALOs(student) in one ALOs(classroom) 
\end{algorithmic}
\label{classroom_user_prompt}
\end{Prompt}

\begin{Prompt}
\caption{Prompt for Visualization in Midjourney V5}
\begin{algorithmic}[1]
\State The student is 15 years old, female, of Hispanic ethnicity, from a middle socioeconomic status background, and is proficient in both English and Spanish. She has B grades, a 95 percents attendance rate, scores 80 percents on tests, completes 90 percents of her homework, and actively participates in class. She is proficient in critical thinking and communication, developing in problem-solving and creativity, and advanced in collaboration. Her interests include math, painting, soccer, and she aspires to become an architect. She prefers visual learning methods. In terms of social behavior, she has friendly peer relations, good conflict resolution skills, emerging leadership abilities, high empathy, and cultural awareness. --v 5. 
\State John Smith is a 35-year-old male teacher with contact information: johnsmith@email.com and +1-555-123-4567. He has a Master's degree in Education, a State Teaching License, and expertise in mathematics. With 10 years of teaching experience in public schools, John has taught mathematics to students in grades 6-8. He is an expert in subject knowledge and communication, and proficient in teaching methods, classroom management, and assessment. John's teaching style is student-centered, inquiry-based, collaborative, and incorporates blended learning, while occasionally using direct instruction. --v 5 
\State In this classroom, Jane Doe is the science teacher with an inquiry-based teaching style and expert communication skills. The class has a total of 25 diverse students with mixed academic performance, developing skills, and varied interests. The classroom is furnished with 25 tables and chairs, a whiteboard, a projector, and 2 shelves. Learning materials include science textbooks, lined notebooks, pens and pencils for stationery, digital resources such as online videos, and lab equipment like microscopes. The environment is characterized by adequate lighting, a comfortable temperature of 22°C, low noise levels, good air quality, and a well-maintained level of cleanliness. --v 5
\State Jane Doe starts the Science lesson using an InquiryBased teaching style. She engages the students with a thought-provoking question related to the topic of the day. The students are asked to collaborate in groups to discuss their ideas and hypotheses. As the students discuss, Jane Doe walks around the classroom, listening to their conversations and offering guidance when needed. She uses the whiteboard and projector to display relevant information and visual aids to support the students' learning. During the lesson, Jane Doe monitors the students' progress and adjusts her teaching methods accordingly. She identifies students who are struggling with the topic and provides individual support to ensure their understanding. At the end of the lesson, Jane Doe assesses the students' comprehension of the topic through a short quiz or group presentation. She then collects feedback from the students to evaluate the effectiveness of her teaching methods and identify any areas for improvement. --v 5 
\end{algorithmic}
\label{alo_jsuser}
\end{Prompt}

\subsection{Case Study 3: IoT scenarios}
 The chosen scenario involves establishing a connection between a smartphone and a printer via a Wi-Fi router in an Internet of Things (IoT) environment. The objective of this case study is to demonstrate that ALOs can often create information too specific about the performance of the system, but not the visual side of system. 
\begin{figure*}[t]
\includegraphics[width=\textwidth]{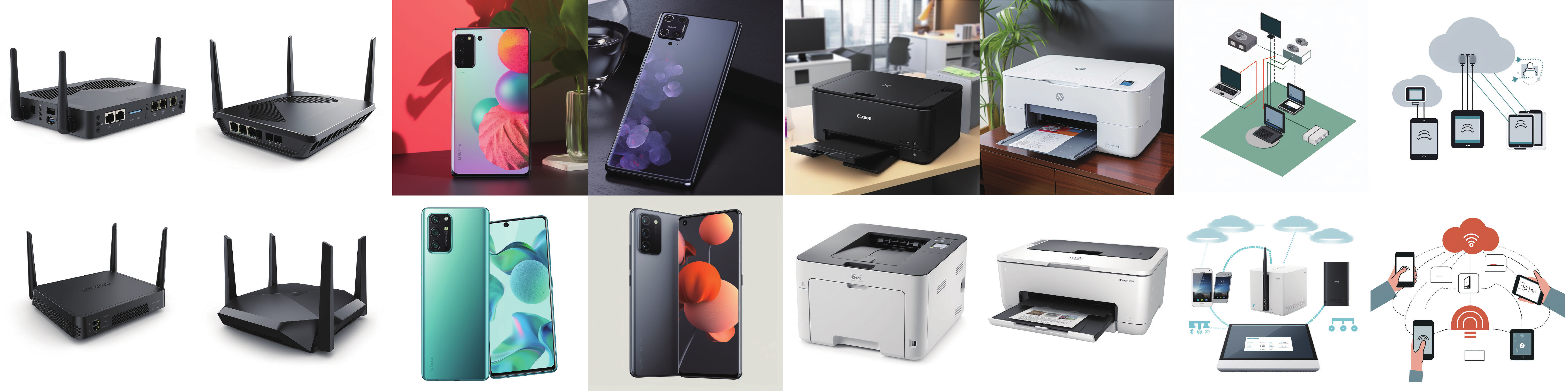}
\caption{Visualization of ALOs(WiFi Router), ALOs(Smartphone), ALOs(printer), and ALOs(interaction) in Midjourney V5 by Prompt 7}
\centering
\label{iot_example}
\end{figure*}
\subsubsection{Contextual Assumptions}
For the purpose of this case study, we make the following contextual assumptions:

1. The IoT devices involved include a smartphone, a printer, and a Wi-Fi router.
2. The devices are connected through a Wi-Fi network, allowing seamless communication between them.
3. Device specifications and parameters are available in the form of tables.
4. Image generation software is used to create visuals for the digital objects based on the provided specifications.

\subsubsection{Connecting a Smartphone and a Printer}
The objective of this step is to use ALOs to perform a language simulation of an IoT device in which a smartphone and a printer are connected via Wi-Fi. To achieve this, we create three ALOs: ALOs (smartphone), ALOs (printer), and ALOs (Wi-Fi router), as seen in Prompt~\ref{phone_printer_prompt}.

\begin{Prompt}
\caption{User Prompt: ALOs Object Creation for WiFi Router, Smartphone and Printer}
\begin{algorithmic}[1]
\State ALOs(WiFi router) and brainstorm all parameters step-by-step to add and fill.
\State get ALOs(WiFi router) object and brainstorm to fill subobject parameters and output one ALOs(classroom) object subobject list and parameters in table
\State ALOs(Smartphone) and brainstorm all parameters step-by-step to add and fill.
\State get ALOs(Smartphone) object and brainstorm to fill subobject parameters and output one ALOs(student) object subobject list and parameters in table
\State ALOs(Printer) and brainstorm all parameters step-by-step to add and fill.
\State get ALOs(Printer) object and brainstorm to fill subobject parameters and output one ALOs(teacher) object subobject list and parameters in table
\State simulate ALOs(teacher) teaches  25 ALOs(student) in one ALOs(classroom) 
\end{algorithmic}
\label{phone_printer_prompt}
\end{Prompt}

First, we define the classes for the smartphone, printer, and Wi-Fi router, taking into account their respective specifications and parameters. Next, we create instances of these classes and establish connections between them using appropriate methods and properties. This simulates the IoT environment where the devices communicate with each other through the Wi-Fi network.

\subsubsection{Creating Visuals using Midjourney V5}
In this step, we employ text2image (Midjourney V5) to create visuals for the digital objects (i.e., smartphone, printer, and Wi-Fi router) based on their specifications, as described in 4.2.1. As mentioned earlier, in many cases, such as product parameters, the specification information is initially filled in, and the parameters related to the visuals such like colors, images on LCDs, photography in real world use cases are not generated as shown in Figure~\ref{iot_example}. 

\begin{Prompt}
\caption{Prompt for Visualization in Midjourney V5}
\begin{algorithmic}[1]
\State The smartphone features a 6.1-inch AMOLED display with a resolution of 1080x2400 and a 60Hz refresh rate. It is powered by a Snapdragon 888 chipset with an octa-core CPU, Adreno 660 GPU, and has high performance. The battery has a 4500mAh capacity, supports 65W charging speed, lasts 1.5 days, and has wireless charging capabilities. The rear camera setup includes a 64MP, 12MP, and 5MP configuration, and the front camera is 32MP. The device records 4K video at 30fps and has features like Night Mode, Portrait, and HDR. It runs on Android 12 with 3 years of updates and regular security patches. The smartphone supports WiFi 6, Bluetooth 5.1, GPS, and 5G networks. It has 128GB of internal storage (non-expandable) and 8GB of RAM. Sensors include an in-display fingerprint sensor, FaceID, gyroscope, proximity, and ambient light. The phone measures 160 x 74 x 8mm, weighs 190g, is made of glass, and has an IP68 water resistance rating. --v 5
\State The router features a 1.4GHz dual-core processor, 512MB of RAM, and 128MB of flash memory. It supports WiFi 6 (802.11ax) with 2.4GHz and 5GHz frequency bands, 4x4 MIMO, and WPA3 encryption. The router has a SPI/NAT firewall and parental controls for security. It offers 1 WAN, 4 LAN, and 2 USB ports (1x USB 2.0, 1x USB 3.0). The coverage area is 3000 sq ft, with 4 external antennas and beamforming technology. The maximum data rate is 6000 Mbps, and the bandwidth is 160MHz. The router supports guest networks, QoS, and VPN features. Setup is accessible through a web interface, app control, or WPS. The router measures 10 x 7 x 2 inches, weighs 1.5 lbs, and is made of plastic material. --v 5
\State The inkjet printer offers color printing with a resolution of 4800x1200 dpi for color and 1200x1200 dpi for mono. It has a color print speed of 10 ppm and a mono print speed of 15 ppm. Connectivity options include USB, WiFi, Ethernet, and mobile printing. The printer has a 150-sheet input capacity and a 50-sheet output capacity. It supports various paper sizes, including A4, A5, A6, B5, Letter, and Legal, as well as automatic duplex printing. The LCD display is 2.7 inches. The printer uses individual ink cartridges and has a yield of 350 pages for color and 400 pages for mono. It features a flatbed scanner with an optical resolution of 1200x2400 dpi and a scan speed of 8 ppm. The printer measures 17.3 inches in width, 13.8 inches in depth, 6.3 inches in height, and weighs 14.3 lbs. --v 5
\end{algorithmic}
\label{prompt_iot_text2img}
\end{Prompt}

By inputting the specifications and parameters into the image generation software, we generate accurate visuals for the digital objects involved in the IoT scenario. This further enhances the language simulation, making it more realistic and comprehensive.

\section{Discussions}

\subsection{Variability of the Response}
Since OpenAI's API is used in this study, the process itself is a black-box, and the response from the LLMs could vary. Furthermore, it is quantitatively difficult to compare the output when the responses are in the natural language or computational code. Thus, we applied embeddings to quantitatively compare the relatedness of each responses. Embeddings are commonly used in searches , clustering and diversity measurements to identify the most similar texts messages. 
\begin{figure*}[t]
\includegraphics[width=0.9\textwidth]{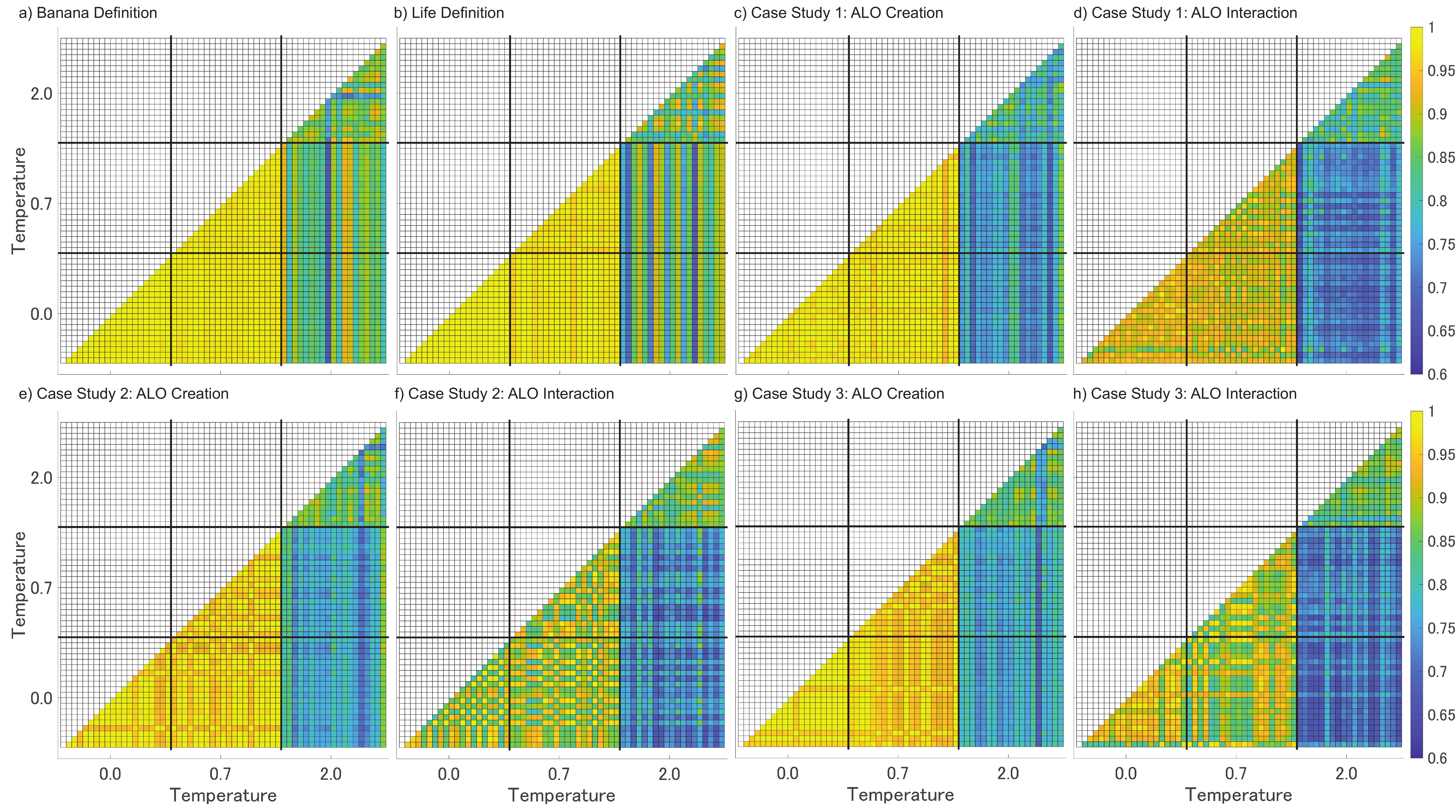}
\caption{Cosine similarity matrix comparing the similarity between each prompt response using embedding (text-embedding-ada-002). The upper half of the matrix is blanked out as it is duplicate image of lower half. Each pixel indicates the prompt response, and there are twenty sets of responses for each temperature (values between 0-2, where 0 is least random and 
2 is most random in GPT-4 API). (a) and (b) are test case where GPT-4 is tasked to define banana and meaning of life, respectively. (c), (e) and (g) shows the cosine similarity matrix of LLM's response when prompted to create ALOs, and (d), (f), and (h) shows the cosine similarity matrix of LLM's response when prompted to interact ALOs with each other.}
\centering
\label{cosine_similarity}
\end{figure*}
Here, we employ text2vector embedding API from OpenAI (text-embedding-ada-002), and forwarded the responses back from the LLM to the embedding API. The API responds with a vector of size 1 by 1536. For example, when GPT-4 is prompted to ``Define the meaning of life in 300 words.'' or ``Define banana in 300 words.''; the mean cosine similarity ($
\text{similarity}(\textbf{a}, \textbf{b}) = \frac{\textbf{a} \cdot \textbf{b}}{|\textbf{a}| |\textbf{b}|}$) of responses (N=20, temperature = 0.0) are 0.988 (S.D. 0.00343) and 0.982 (0.00482), respectively (see Figure~\ref{cosine_similarity} (a)-(b)). Temperature settings of the API determines the randomness of the response with 0 being most focused and deterministic, while 2 outputs more random response. By increasing the temperature of the API to 2.0, the mean cosine similarity of responses drops to 0.848 (S.D. 0.0528) and 0.859(S.D. 0.0495), respectively. The default temperature used in this study is 0.7, and the responses are relatively consistent with cosine similarity being 0.971 (S.D. 0.00819) and 0.975 (S.D. 0.00505) for banana and life definitions, respectively. This demonstrates that the temperature of 0.7 consistently returns similar response in comparison to temperature setting of 2.0. 

We performed similarity analysis to all the case studies in section 4, and the results are as shown in Figure~\ref{cosine_similarity} (c)-(h). Here, we simplified the task to allow for easier comparison across the scenarios, and used Prompt~\ref{alo_creation} as the system prompts (specific examples with Javascript code generation is separately made available in the supplementary material). The ALO creation prompt were ``ALOs(universe), ALOs(cat), ALOs(roomba(robot creaner))'', ``ALOs(class room), ALOs(student), ALOs(teacher)'' and ``ALOs(wifi router), ALOs(smartphone), ALOs(printer)'' for case study 1, 2 and 3, respectively. After crating ALO object for the each case study, the ALO objects were tasked to interact with each other. The user prompt for interaction was ``ALOs cat meet ALOs roomba(robot cleaner)'', ``ALOs teacher teaches ALOs student'', and ``ALOs smartphone connects to ALOs printer'' for case study 1, 2 and 3. 

Taking home environment case studies as an example, the cosine similarity index for temperature 0.0, 0.7 and 2.0 in ALOs creation prompt are 0.978 (S.D. 0.00975), 0.973 (S.D. 0.0161), 0.804 (S.D. 0.0436), respectively (Figure~\ref{cosine_similarity} (c)). This demonstrates that the LLMs consistently create similar ALOs object, and this trends are also continue to be seen in the other case studies in Figure~\ref{cosine_similarity} (c) and (e). The mean cosine similarity index for case study 2 (Figure~\ref{cosine_similarity} (c)) are 0.967 (S.D. 0.0227), 0.963 (S.D. 0.0182), and 0.837 (S.D. 0.0484) for temperature 0, 0.7 and 2, respectively. Similarly, the mean cosine similarity index for case study 3 (Figure~\ref{cosine_similarity} (e)) are 0.980 (S.D. 0.0160), 0.953 (S.D. 0.0191), 0.837 (S.D. 0.0454) for termperature 0.0, 0.7 and 2.0, respectively. 

The variability in the responses are most prevalent when the ALOs are asked to interact with each other as shown in Figure~\ref{cosine_similarity} (d), (f) and (h). The mean cosine similarity index drops to 0.922 (S.D. 0.026), 0.909 (S.D. 0.0307), and 0.834 (S.D. 0.0.86) for temperature 0.0, 0.7 and 2.0, respectively. This trends continue for case study 2, and the mean cosine similarity index is 0.895 (S.D. 0.0595), 0.900 (S.D. 0.0512), and 0.861 (0.0367) for temperature 0.0, 0.7 and 2.0 (Figure~\ref{cosine_similarity} (f)). For case study 3, the mean cosine similarity index is 0.920 (S.D. 0.0473), 0.898 (S.D. 0.0469), and 0.847 (S.D. 0.0334) for termperature 0.0, 0.7 and 2.0, respectively (Figure~\ref{cosine_similarity} (h)). The most demanding aspect of domain expertise for both LLMs and users arises when variability occurs during the interaction stage. Users must ensure that the ALO possesses sufficient properties and functions to perform as intended. Nevertheless, if any functions or properties are missing in the ALOs, users can simply supplement the necessary information in subsequent prompts. The remarkable ability of LLMs to connect ALOs, even in ambiguous situations, exemplifies their robustness and versatility. Furthermore, the capacity of LLMs to diverge and adapt contributes significantly to bridging the gap between TMO and LO, effectively filling in any missing pieces and enhancing their overall performance.

\subsection{Comparison of Abstraction Levels}

\subsubsection{ALOs vs. Prompts}
ALOs and Prompts are similar in that they both serve as input to computational models. However, ALOs provide a higher level of abstraction, allowing for more complex interactions and relationships between objects. Prompts, on the other hand, serve as simple instructions to guide the model's response. The comparison between ALOs and Prompts is analogous to the difference between OOP and procedural programming languages like C. ALOs provide a more structured and organized approach, while Prompts offer a more direct and straightforward means of communication.

\subsubsection{ALOs vs. Linking to TMOs}
ALOs and TMOs represent different aspects of computational linguistics. ALOs focus on the linguistic abstraction of objects, whereas TMOs deal with their computational implementation. 

\subsubsection{ALOs vs. LOs}
ALOs and LOs are closely related as they both represent linguistic abstractions of objects. However, ALOs provide a higher level of abstraction than LOs, allowing for more complex object interactions and relationships. In this sense, ALOs can be considered an extension of LOs, with the potential to facilitate more advanced object-oriented descriptions and interactions.

\subsubsection{LOs to TMOs}
The connection between LOs and TMOs is crucial for realizing the Digital Nature vision. By linking LOs to TMOs via ALOs, we can bridge the gap between human perception and computational processes, enabling seamless interaction between real-world objects and their digital simulations. However, this linkage can be challenging due to the inherent differences in their representations and the need for efficient translation mechanisms with ALOs like intermediate LOs.

\subsubsection{Parameter Accessibility}
Parameter accessibility is an essential aspect of object representations which enable users to manipulate and modify object properties easily. Prompts and LOs provide a higher level of abstraction, which can sometimes hinder direct parameter access. On the other hand, TMOs and ALOs offer more direct and straightforward means of interaction, potentially facilitating greater parameter accessibility.

\subsection{Security Risks and Concerns}
Our approach raises several security-related concerns that warrant further investigation by the HCI community. For instance, as LLMMs become more adept at understanding and manipulating digital objects, they may also become more susceptible to adversarial attacks, where malicious actors exploit vulnerabilities in the models to compromise their functionality or integrity.
Additionally, the automated transformation of real-world objects into digital twins could potentially introduce new attack vectors, as adversaries may seek to manipulate the digital representations of critical physical systems (e.g., power grids, transportation networks) to cause harm or disruption. To mitigate these risks, it is crucial for the HCI community to develop robust security mechanisms and protocols that can protect both the underlying LLMMs and the digital twins generated by our approach.
\subsection{Limitations}

\subsubsection{Domain Knowledge}
The absence of domain knowledge in ALOs leads to unpredictable behavior as the underlying parameters that govern the real-world phenomena are not incorporated within the ALOs. This unpredictability poses a significant challenge in effectively simulating and manipulating real-world objects within LLMMs and high-level programming languages. Consequently, the seamless interaction between digital and physical realities, as envisioned in Digital Nature, remains unattainable.
To address this issue, we propose integrating domain knowledge into ALOs, thereby enhancing their predictability and facilitating more accurate simulations. This integration allows ALOs to not only represent real-world objects but also to incorporate the rules and constraints that govern their behavior. 

\subsubsection{Token Length}
The size of object tokens is a critical factor that determines the efficiency of ALOs. Compression techniques play a significant role in minimizing the token size, thereby reducing the computational resources required for processing and storage. Various research has suggested method of connecting LLM to other databases to extend LLM's memory\cite{Tshitoyan2019} and thus reduce the token size\footnote{https://github.com/mayooear/gpt4-pdf-chatbot-langchain}. ALOs can similarly extend its capability by storing and registering the ALO objects in database and making it available for reference. 
\section{Conclusion and Future Directions}
In this paper, we present a new approach called Abstract Language Objects (ALOs), which simplifies the relationship between Linguistic Objects (LOs) and Turing Machine Objects (TMOs) in object-oriented programming (OOP) and language simulation. ALO recursively updates each object as new ecosystems are organized with each new species discovered in the complex world of taxonomy. By leveraging the abstraction capabilities of Large Language Models (LLMs), ALOs facilitate seamless communication between human-understood real-world entities and their computational equivalents. This method effectively implements and manages real-world simulations, ultimately advancing the concept of Digital Nature.

\section{Acknowledgments}
The manuscript was drafted using deepL write, OpenAI ChatGPT and GPT-4. The AI generated text was read, revised and proofed by the authors. We thank Takahito Murakami for drawings.
\vspace{-10pt}
\bibliographystyle{ACM-Reference-Format}
\bibliography{sample-base}


\end{document}